\documentclass[final,3p,times]{elsarticle}

\usepackage{graphicx}
\DeclareGraphicsExtensions{.eps}

\usepackage{algorithm}
\usepackage{algpseudocode}
\algblock[MyInitBlock]{StartInit}{EndInit}
\algrenewtext{StartInit}{\textbf{Initialize:}}%
\algnotext[MyInitBlock]{EndInit}

\algblock[MyDefineBlock]{StartDefine}{EndDefine}
\algrenewtext{StartDefine}{\textbf{Define:}}%
\algnotext[MyDefineBlock]{EndDefine}

\algnewcommand\algorithmicto{to}
\algrenewtext{For}[3]%
{\algorithmicfor\ $ #1 \gets #2$ \algorithmicto\ $#3$ \algorithmicdo}

\algsetblock[MyInitBlock]{StartInit}{}{}{0.2cm}
\algsetblock[MyDefineBlock]{StartDefine}{}{}{0.2cm}

\algsetblock[]{While}{EndWhile}{}{0.2cm}
\algsetblock[]{For}{EndFor}{}{0.2cm}
\algsetblock[]{If}{EndIf}{}{0.2cm}

\usepackage{amsmath}

\usepackage{amssymb}
\DeclareMathOperator*{\argmin}{arg\,min}
\DeclareMathOperator*{\argmax}{arg\,max}

\usepackage{array}
\usepackage{multirow}

\newcolumntype{x}[1]{%
>{\centering\hspace{0pt}}p{#1}}%

\journal{Digital Signal Processing}

\begin{document}
\begin{frontmatter}

\cortext[cor]{Corresponding author}

\title{A* Orthogonal Matching Pursuit: Best-First Search for Compressed Sensing Signal Recovery}
\author[sabanci,bte]{Nazim Burak Karahanoglu\corref{cor}}
\ead{karahanoglu@sabanciuniv.edu}

\author[sabanci]{Hakan~Erdogan}
\ead{herdogan@sabanciuniv.edu}

\address[sabanci]{Department of Electronics Engineering, Sabanci University, Istanbul 34956, Turkey}
\address[bte]{Information Technologies Institute, TUBITAK-BILGEM, Kocaeli 41470, Turkey}

\begin{abstract}
Compressed sensing is a developing field aiming at reconstruction of sparse signals acquired in reduced dimensions, which make the recovery process under-determined. The required solution is the one with minimum $\ell_0$ norm due to sparsity, however it is not practical to solve the $\ell_0$ minimization problem. Commonly used techniques include $\ell_1$ minimization, such as Basis Pursuit (BP) and greedy pursuit algorithms such as Orthogonal Matching Pursuit (OMP) and Subspace Pursuit (SP). This manuscript proposes a novel semi-greedy recovery approach, namely A* Orthogonal Matching Pursuit (A*OMP). A*OMP performs A* search to look for the sparsest solution on a tree whose paths grow similar to the Orthogonal Matching Pursuit (OMP) algorithm. Paths on the tree are evaluated according to a cost function, which should compensate for different path lengths. For this purpose, three different auxiliary structures are defined, including novel dynamic ones. A*OMP also incorporates pruning techniques which enable practical applications of the algorithm. Moreover, the adjustable search parameters provide means for a complexity-accuracy trade-off. We demonstrate the reconstruction ability of the proposed scheme on both synthetically generated data and images using Gaussian and Bernoulli observation matrices, where A*OMP yields less reconstruction error and higher exact recovery frequency than BP, OMP and SP. Results also indicate that novel dynamic cost functions provide improved results as compared to a conventional choice.

\end{abstract}


\end{frontmatter}

\section{Introduction \label{Sec:Intro}}
Compressed sensing (CS) deals with the acquisition of the sparse signals, i.e. signals with only a few nonzero coefficients, in reduced dimensions. As a natural consequence of this, the signal has to be reconstructed back to its full dimension using the observation in reduced dimensions. CS is based on the following question: Can a reduced number of observations (less than Shannon-Nyquist rate) contain enough information for exact reconstruction of sparse signals? One might argue that this seems quite unnatural, however a number of articles in CS literature, i.e. \cite{Candes:RUP}, \cite{Donoho:CS} and \cite{Candes:NOptRec}, state that it is indeed possible under certain assumptions.

Exact solution of the CS reconstruction problem requires minimization of the $\ell_0$ norm, i.e. the number of nonzero coefficients, which is unpractical. One of the solutions that can be found in the literature is the convex relaxation which replaces $\ell_0$ minimization problem with an $\ell_1$ minimization, such as Basis Pursuit \cite{Chen:BP}. Another family of algorithms, so called greedy pursuit algorithms, Orthogonal Matching Pursuit (OMP) \cite{Pati:OMP}, Subspace Pursuit (SP) \cite{Dai:SP}, Iterative Hard Thresholding (IHT) \cite{Blumensath:IHT1,Blumensath:IHT2} etc. provide greed and find approximate solutions by solving a stagewise constrained residue minimization problem.

This manuscript proposes a new semi-greedy CS reconstruction approach that incorporates the A* Search \cite{Koenig:AStar, Dechter:AStar, Jelinek:SMSP, Hart:FBHDMCP, Hart:CorrFBHDMCP}, a best-first search technique that is frequently used in path finding, graph traversal and speech recognition. This new method, which we call A*OMP, proposes an A* search that employs the OMP algorithm to expand the most promising path of the search tree at each iteration. By utilizing best-first search, multiple paths can be evaluated during the search, which promises improvements over the single path structures of algorithms such as MP or OMP. This combination of A* search and OMP is not straightforward: It requires appropriately defined cost models which enable A* to perform stage-wise residue minimization in an intelligent manner, and effective pruning techniques which make the algorithm tractable in practice. As for the cost model, which should make comparison of paths with different lengths possible, we introduce two novel dynamic structures, which better comply with our needs, in addition to the trivial additive one. Pruning capability is provided via a number of strategies which, together with the cost model parameters, enable a complexity-accuracy trade-off. The effectiveness of the proposed pruning techniques and the dynamic cost models is demonstrated via provided reconstruction examples. This reconstruction experiments, including different nonzero coefficient distributions, Gaussian and Bernoulli type random observation matrices, noise contaminated measurements and images, demonstrate that utilization of best-first search is able to improve the reconstruction accuracy. A preliminary version of this work has been presented in \cite{AOMP_ICASSP}.

A number of tree-search based methods have appeared in CS literature. These methods are, however, fundamentally different than A*OMP as they do not follow the best-first search principle. The tree-search based OMP (TB-OMP), \cite{Cotter:TSBOMP}, employs a tree-search that opens $L$ children per each node at a level. A rather flexible version of this is the flexible tree-search based OMP (FTB-OMP) \cite{Karabulut:FlexTreeSearch}, where the branching factor $L$ is decreased at each level. Another straightforward tree-search also appears in Fast Bayesian Matching Pursuit \cite{Schnitter:FBMP}, which opens all children of the nodes at a level, and retains the best $D$ wrt. their posterior probabilities. These methods incorporate rather simple and non-sophisticated tree-search techniques in comparison to A*OMP. They employ neither cost models to compensate for different path lengths, nor mechanisms to select the most promising path on the fly, but expand all nodes at a level. They do not also possess effective pruning techniques, except FTB-OMP pruning the children of a node wrt. their correlations to the best one, and FBMP keeping $D$ nodes at a level. The randomized OMP (RandOMP) algorithm \cite{Elad:PluralSparRep} yields an estimate of the minimum mean-squared error (MMSE) solution by averaging multiple sparse representations which are obtained by running a randomized version of OMP several times. Though RandOMP involves multiple sparse representations, it incorporates no explicit tree-search.

To avoid some possible misunderstanding, we would like to note that the tree search concept in A*OMP is completely general to all sparse signals. A*OMP aims to find a closer result to the true $\ell_0$ solution, thus the objective is to improve reconstruction quality not to decrease computational complexity to find a greedy solution, such as in list decoding \cite{Phan:ListDecoding}. Furthermore, A*OMP is neither specific for tree-sparse signals nor does it make use of a tree-structured over-complete basis as for the tree-based OMP algorithm \cite{La:TreeOMP}. The algorithm is not specific for structured sparse signals as well.

The rest of this manuscript is organized as follows: CS reconstruction problem and some major algorithms are introduced briefly in sections~\ref{Sec:CS}. A* search is discussed in section~\ref{Sec:AStar}. Section~\ref{Sec:SparRecAStar} is devoted to the A*OMP algorithm and the novel cost functions. We demonstrate the reconstruction performance of A*OMP in comparison to Basis Pursuit (BP) \cite{Chen:BP}, Subspace Pursuit (SP) \cite{Dai:SP} and OMP \cite{Pati:OMP} in Section~\ref{Sec:Results}, before concluding the manuscript with a short summary.

\section{Compressed Sensing \label{Sec:CS}}

\subsection{Problem Definition \label{Sec:Def}}
Compressed Sensing acquisition of a $K$-sparse signal $\mathbf{x}$, i.e. having only $K$ nonzero entries, is obtained via the observation matrix, or dictionary, $\mathbf{\Phi}$:
\begin{equation}
\label{Eq:MeasBasisRep}
    \mathbf{y}=\mathbf{\Phi}\mathbf{x}
\end{equation}
where $\mathbf{x}\in{\mathbb{R}}^{N}$, $\mathbf{\Phi}\in{\mathbb{R}}^{M\times{N}}$, $\mathbf{y}\in{\mathbb{R}}^{M}$ and $K<M<N$. As $M<N$, solving for $\mathbf{x}$ directly from (\ref{Eq:MeasBasisRep}) is ill-posed. CS exploits sparsity of $\mathbf{x}$ to formulate the reconstruction problem alternatively as
\begin{equation}\label{Eq:L0Minimization}
\mathbf{x}=\argmin\|\mathbf{x}\|_{0} \;\;\; s.t. \;\;\; \mathbf{y}=\mathbf{\Phi}\mathbf{x}.
\end{equation}
where $\|.\|_{0}$ denotes the $\ell_0$ norm, which is the number of nonzero coefficients of a signal. Solving (\ref{Eq:L0Minimization}) directly is not feasible as it requires an exhaustive combinatorial search \cite{Candes:RUP, Candes:ErrCorr}. Consequently, a variety of strategies have emerged to find approximate solutions to (\ref{Eq:L0Minimization}).

\subsection{Theoretical Guarantees - The Restricted Isometry Property  \label{Sec:RIP}}

An important means for obtaining theoretical guarantees in CS recovery problem is the restricted isometry property (RIP) \cite{Candes:DecLP, Candes:NOptRec, Candes:RIP}: A matrix $\mathbf{\Phi}$ is said to satisfy the $K$-RIP if there exists a restricted isometry constant $\delta_K$, $0<\delta_K<1$ such that
\begin{equation}\label{Eq:RIP}
    (1-\delta_K)\|\mathbf{x}\|_2^2 \leq \|\mathbf{\Phi}\mathbf{x}\|_2^2 \leq (1+\delta_K)\|\mathbf{x}\|_2^2, \forall \mathbf{x}: \|\mathbf{x}\|_0 \leq K .
\end{equation}
A matrix satisfying the RIP acts almost like an orthonormal system for sparse linear combinations of its columns \cite{Candes:DecLP}, making reconstruction of sparse signals from lower dimensional observations possible.

Analysis in \cite{Candes:NOptRec, Rudelson:SparseRec} state that matrices with i.i.d. Gaussian or Bernoulli entries and matrices randomly selected from discrete Fourier transform satisfy the RIP with high probabilities, when they satisfy some specific conditions on $K$ based on $M$ and $N$. Therefore, such random observation matrices can provide compact representations of sparse signals.

\subsection{Major CS Reconstruction Algorithms\label{Sec:MajorFam}}

Following \cite{Tropp:CompMeth}, CS recovery approaches can be categorized as greedy pursuit algorithms, convex relaxation, Bayesian framework, nonconvex optimization and brute force methods. In this work, we are interested in the first two of these.

\subsubsection{Convex Relaxation \label{Sec:BP}}

$\ell_1$ or convex relaxation algorithms rely on the relaxation of the $\ell_0$ norm minimization in (\ref{Eq:L0Minimization}) by an $\ell_1$ norm, which first appeared in Basis Pursuit \cite{Chen:BP}. In this context, (\ref{Eq:L0Minimization}) is rewritten as
\begin{equation}\label{Eq:L1Minimization}
\mathbf{x} = \argmin\|\mathbf{x}\|_{1} \;\;\;   s.t \;\;\; \mathbf{y}=\mathbf{\Phi}\mathbf{x},
\end{equation}
which can be solved via computationally tractable convex optimization methods, such as pivoting, linear programming and gradient methods \cite{Tropp:CompMeth}. Extensive analysis of RIP conditions for $\ell_1$ relaxation can be found in \cite{Candes:DecLP, Candes:NOptRec, Candes:RIP, Foucart:SparsestSolLq}.

\subsubsection{Greedy Pursuits \label{Sec:Greedy}}

Historically, Matching Pursuit (MP) \cite{Mallat:MP} is the first greedy pursuit. MP expands the support of $\mathbf{x}$ by the dictionary atom which has the highest inner-product with the residue at each iteration. Major drawback of MP is that it does not take into account the non-orthogonality of the dictionary, which results in suboptimal choices of the nonzero coefficients.

The non-orthogonality of dictionary atoms is taken into account by the Orthogonal Matching Pursuit (OMP) \cite{Pati:OMP}, which performs orthogonal projection of the residue onto the selected dictionary atoms after each iteration. Ensuring orthogonality of the residue to the selected support enhances the reconstruction. As expansion of paths in A*OMP is very similar to OMP, we devote some space to a short overview of this method.

Let's first define the notation: Let $v_n\in{\mathbb{R}}^{M},n=1,2,...,N$ be the dictionary atoms, i.e. columns of the dictionary $\mathbf{\Phi}$. $\mathbf{r}^l$ denotes the residue after the $l$'th iteration. $\mathbf{S}$ and $\mathbf{c}$ denote the matrix (or, exchangeably in context, the set) of atoms selected from $\mathbf{\Phi}$ for representing $\mathbf{y}$ and the vector of corresponding coefficients respectively.

OMP is initialized as $\mathbf{r}^0=\mathbf{y}$, $\mathbf{S}=\{\}$ and $\mathbf{c} = \mathbf{0}$. At iteration $l$, OMP appends $\mathbf{S}$ the dictionary atom that best matches $\mathbf{r}^{l-1}$
\begin{eqnarray}
\label{Eq:MPSelectStep}
    \mathbf{s} &=& \argmax_{\mathbf{v}_n\in{\mathbf{\Phi}\setminus{\mathbf{S}}}}\langle\mathbf{r}^{l-1},\mathbf{v}_n\rangle,\nonumber \\
    \mathbf{S} &=& \mathbf{S} \cup \mathbf{s}.
\end{eqnarray}
The coefficients are computed by the orthogonal projection
\begin{equation}
\mathbf{c} = \argmin_{\mathbf{\tilde{c}}\in{\mathbb{R}}^{l}} \|\ \mathbf{y} - \mathbf{S}\mathbf{\tilde{c}} \|_{2}.
\end{equation}
At the end of each iteration, the residue is updated as
\begin{equation}
\label{Eq:MPUpdateRes}
    \mathbf{r}^l = \mathbf{y} - \mathbf{S}\mathbf{c}.
\end{equation}

After termination, $\mathbf{S}$ and $\mathbf{c}$ contain the support and the corresponding nonzero entries of $\mathbf{x}$, respectively. OMP may employ different termination criterion. In this work, we fix the number of iterations as $K$. Alternatively, iterations can be carried on until the residue falls below a threshold.

A detailed analysis of OMP is provided in \cite{Tropp:OMP} which states a lower-bound on the number of observations for exact recovery. The guarantees for OMP, however, were shown to be non-uniform, i.e. they hold only for each fixed sparse signal, but not for all  \cite{Needell:ROMP}. It was shown in \cite{Rauhut:NonUniformGreedy} that for natural random matrices it is not possible to obtain uniform guarantees for OMP.

Recently, more sophisticated pursuit methods, which select multiple columns per iteration, have appeared. For example, Stagewise OMP (StOMP) \cite{Donoho:StOMP} selects in each step all columns whose inner-products with the residue is higher than an adaptive threshold depending on the $\ell_2$ norm of the residue. Alternatively, regularized OMP (ROMP)  \cite{Needell:ROMP} groups inner-products with similar magnitudes into sets at each iteration and selects the set with maximum energy. Via this regularization, ROMP provides RIP-based uniform guarantees. Compressive Sampling Matching Pursuit (CoSaMP) \cite{Needell:CoSaMP} and Subspace Pursuit (SP) \cite{Dai:SP} combine selection of multiple columns per iteration with a pruning step. At each iteration, these first expand the selected support by addition of new atoms, and then prune it to retain only the best K atoms. Both CoSaMP and SP are provided with optimal performance guarantees based on RIP.

Iterative hard thresholding (IHT) \cite{Blumensath:IHT1,Blumensath:IHT2} employs an iterative gradient search that first updates the sparse estimate in the direction of the gradient of the residue wrt. the dictionary and then prunes the solution by either thresholding or keeping only the K largest entries. IHT is equipped with RIP based guarantees similar to CoSaMP and SP \cite{Blumensath:IHT2}. A recent IHT variant, Nesterov iterative hard thresholding (NIHT) \cite{Cevher:NIHT} employs Nesterov's proximal gradient \cite{Nesterov} to update the sparse representation. NIHT provides no a priori performance guarantee, but still an online performance guarantee.

\section{A* Search \label{Sec:AStar}}

A* search \cite{Koenig:AStar, Dechter:AStar, Jelinek:SMSP, Hart:FBHDMCP, Hart:CorrFBHDMCP} is an iterative tree-search algorithm. In our problem, the A* search tree is iteratively built up by nodes which represent the dictionary atoms. Each path from the root to a leaf node denotes a subset of dictionary atoms which is a candidate support for $\mathbf{x}$. A path is called \textit{complete} if it has $K$ nodes, and \textit{partial} if it is shorter. A* search tree is initialized with all possible single-node paths. At each iteration, the most promising path is chosen and all of its children are added to the search tree. Search is terminated when the most promising path is found to be complete.

In our scope, A* search looks for the complete path $\mathbf{p}^K$ which minimizes some evaluation function $g(\mathbf{p}^K)$. As tree paths typically have different lengths, these cannot be compared via an evaluation function which depends on the number of nodes on the path. In order to deal with different path lengths, A* search employs an auxiliary function \cite{Jelinek:SMSP}. For a path $\mathbf{p}^l$ of length $l \leq K$, the auxiliary function $d(\mathbf{p}^l)$ is defined such that $d(\mathbf{p}^K) = 0$ and
\begin{equation}\label{Eq:AStar1}
    d(\mathbf{p}^l) \geq  g(\mathbf{p}^l) - g(\mathbf{p}^l \cup \mathbf{z}^{K-l}), \;\;\; \forall \mathbf{z}^{K-l} ,
\end{equation}
where $\mathbf{z}^{K-l}$ is a sequence of $K-l$ nodes and $\cup$ denotes concatenation. With this definition, $d(\mathbf{p}^l)$ is larger than or equal to the decrement in the evaluation function that any complete extension of the path $\mathbf{p}^l$ could yield.

Now, we define the cost function as
\begin{equation}\label{Eq:AStar2}
    F(\mathbf{p}^l) = g(\mathbf{p}^l) - d(\mathbf{p}^l).
\end{equation}
Let's consider a complete path $\mathbf{p}^K$ and a partial path $\mathbf{\widetilde{p}}^l$ of length $l<K$. Combining (\ref{Eq:AStar1}) and (\ref{Eq:AStar2}), if $ F(\mathbf{p}^K) \leq  F(\mathbf{\widetilde{p}}^l)$, then $g(\mathbf{p}^K) \leq g(\mathbf{\widetilde{p}}^l \cup \mathbf{z}^{K-l})$ for all $\mathbf{z}^{K-l}$, which states that $\mathbf{p}^K$ is better than all possible extensions of $\mathbf{\widetilde{p}}^l$. Hence, it is safe to use the cost function $F(.)$ for selecting the most promising path. Note that, satisfying (\ref{Eq:AStar1}) may either be impossible or unpractical in practice. This issue is discussed when different A*OMP cost models are introduced in Section~\ref{Sec:SelBestPath}.

\section{Sparse Signal Reconstruction using A* Search \label{Sec:SparRecAStar}}

\begin{figure}[!t]
\centerline{\includegraphics[width=0.75\textwidth]{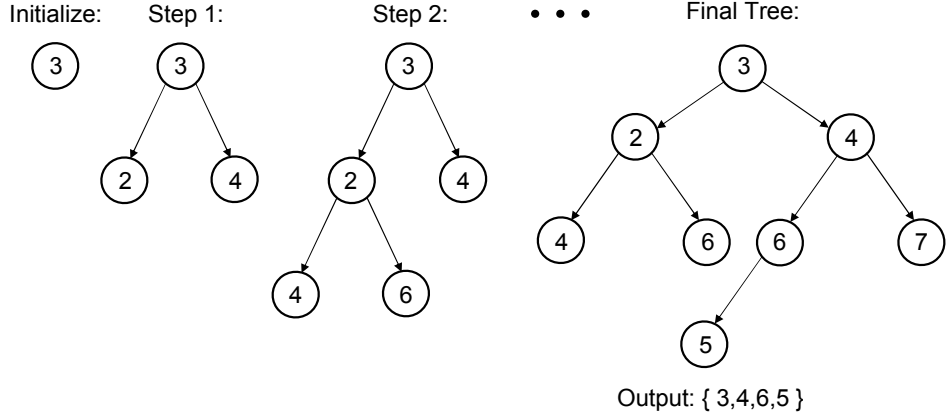}}
\caption{Evaluation of the search tree during A*OMP algorithm}
\label{Fig:AOMP}
\end{figure}

A*OMP casts the sparse recovery problem into a search for the correct support of the $K$-sparse $\mathbf{x}$ among a number of dynamically evolving candidate subsets. These candidate subsets are stored as paths from the root node to leaf nodes of a search tree, where each node represents an atom in $\mathbf{\Phi}$. The search tree is built up and evaluated iteratively by A* search. The search starts with candidate subsets of single elements. At each iteration, new dictionary atoms are appended to the most promising path, which is selected to minimize some cost function based on the residue. In this way, A*OMP performs a multi-path search for the best one among all possible $K$-element subsets of $\mathbf{\Phi}$. Though the A*OMP search tree actually restricts the search to a set of iteratively built candidate subsets, it is general with the capability of representing all possible $K$-element subsets of $\mathbf{\Phi}$. Fig.~\ref{Fig:AOMP} illustrates evaluation of a sample search tree throughout the search.

Incorporation of a multi-path search strategy is motivated by the expectation that it would improve reconstruction especially where a single-path algorithm such as OMP fails because of the linear dependency of dictionary atoms. In cases where computation of a single path yields a wrong representation, the correct one will mostly be in the set of candidate representations. By a properly configured multi-path search, i.e. by proper selection of the cost model as discussed below, this correct path may be distinguished among the candidates. In other words, a multi-path strategy may reduce the error especially when too few measurements are provided.

For the rest of this work, we differentiate between the paths in the search tree with subscripts. The superscripts represent either the length of the path, or the position of the node in the path. $\mathbf{s}^l_i$ represents the selected atom at the $l$'th node on path $\mathbf{S}_i$ and $c^l_i$ the corresponding coefficient. Similarly $\mathbf{r}_i$ is the residue of path $i$. $\mathbf{S}_i$ and $\mathbf{c}_i$ denote the matrix of atoms selected for path $i$ and the vector of corresponding coefficients, respectively. Note that $\mathbf{S}_i$ and $\mathbf{s}^l_i$ are the mathematical equivalents of the corresponding path and node, respectively. In the rest of this work, we slightly abuse this notation and use $\mathbf{s}^l_i$ and $\mathbf{S}_i$ also to represent the corresponding node and path.

We discuss utilization of tree search for A*OMP in three main steps: initialization of the search tree, selecting the best path and expansion of the selected partial path.

\subsection{Initialization of the Search Tree}

A* search originally initializes the search tree by all possible paths with length $1$. This corresponds to $N$ different initial subsets, which is not practical in most cases as $N$ is usually large. In fact, only $K\ll N$ dictionary atoms are relevant to $\mathbf{y}$. Moreover, each iteration adds the tree multiple children of a selected partial path (Section~\ref{Sec:ExpSelPath}). Hence, the search might be started with less paths. As a consequence, we limit the initial search tree to the $I \ll K$ subsets, each of which contains one of the $I$ atoms having the highest absolute inner-product with $\mathbf{y}$. Note that another possibility would be selecting the atoms whose inner-products with $\mathbf{y}$ are greater than a certain threshold.

\subsection{Expanding the Selected Partial Path \label{Sec:ExpSelPath}}

In typical A* search, all children of the most promising partial path are added to the search tree at each iteration. In practice, this results in too many search paths because of the high number of possible children: To illustrate, let the length of the selected partial path be $l$. This path has $N-l \approx N$ children since $l<K \ll N$. Hence, each iteration considers approximately $N$ new paths and the upper bound on the number of paths involved overall in the search is obtained as $N^K$, given $K \ll N$. To limit these, we employ three pruning strategies:

\subsubsection{Extensions per Path Pruning \label{Sec:LimitExtPath}}

For our purposes, order of nodes along a path is unimportant. At each step, we require only to add one of the $K$ correct atoms to the representation, and not a specific one of them.
Therefore, considering only a few children of a selected partial path becomes a reasonable sacrifice. At each A*OMP iteration, we expand the search tree only by the $B$ children which have the highest absolute inner-product with the residue to the selected path. Note that another reasonable choice would be considering only the children whose inner-products with the residue are higher than a threshold.

Extensions per Path Pruning decreases the upper bound on the number of paths from $N^K$ to $B^K$. Starting the search with $I$ initial paths, this bound becomes $I*B^{(K-1)}$. Practically, $I$ and $B$ are chosen much smaller than $N$, decreasing the paths involved in the search drastically.

\subsubsection{Tree Size Pruning \label{Sec:LimitStackSize}}

Despite extensions per path are limited to $B$, adding new paths at each iteration still increases required memory, as the corresponding residues are also necessary. To reduce memory requirements, we adopt the ``beam search'' strategy and we limit the maximum number of paths in the tree by the beam width $P$. When this limit is exceeded, the worst paths, i.e. the ones with maximum cost, are removed from the tree till $P$ paths remain.

\begin{figure*}[!t]
\centerline{\includegraphics{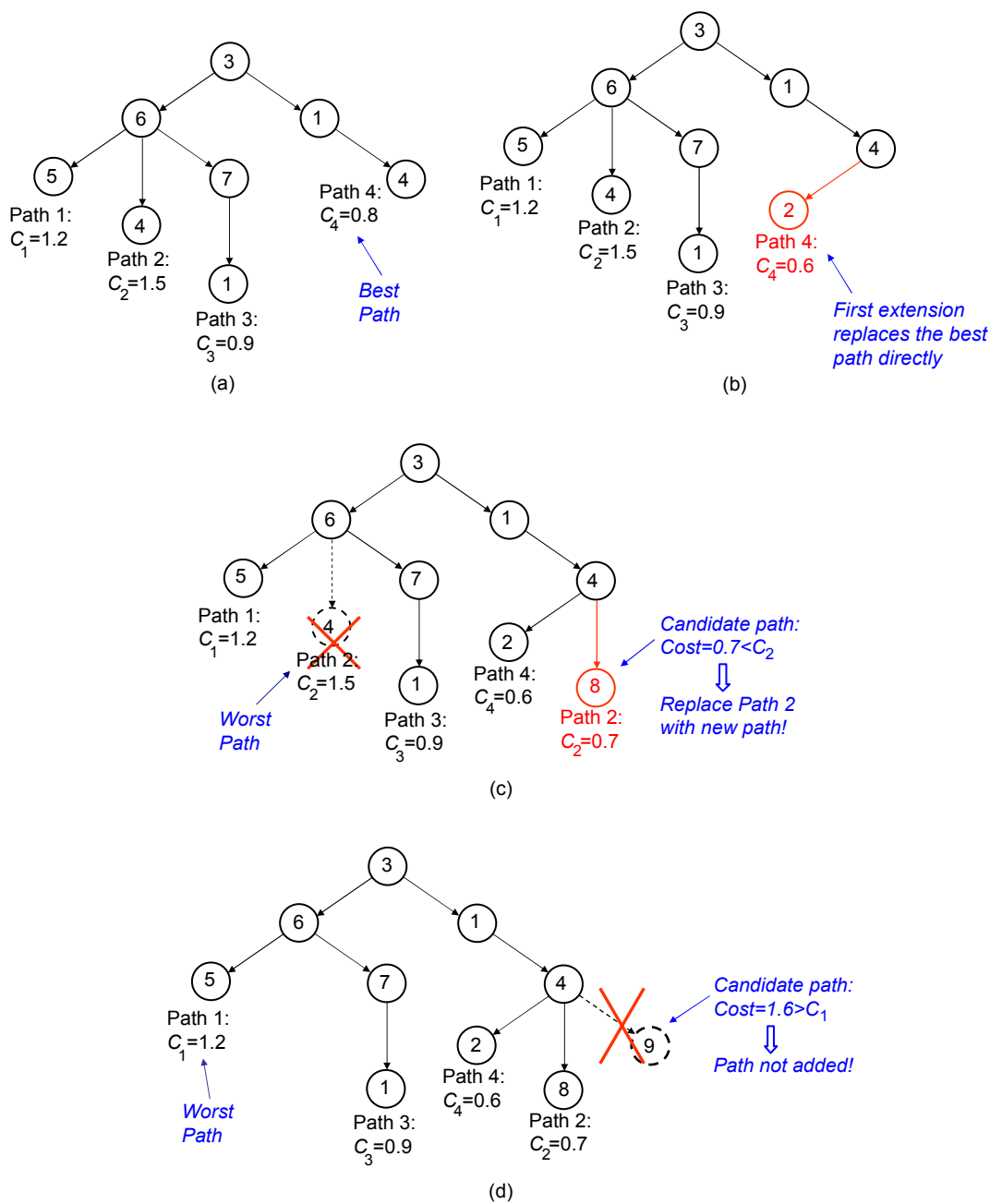}}
\caption{Evaluation of the search tree during a single iteration of the A*OMP algorithm}
\label{Fig:IterationAOMP}
\end{figure*}

Fig.~\ref{Fig:IterationAOMP} illustrates the Extensions per Path and Tree Size Pruning rules where $P=4$ and $B=3$. Fig.~\ref{Fig:IterationAOMP}a depicts a search tree with four paths at the beginning of an iteration. The cost of each path is indicated with $C_i$. Path 4, which has the minimum cost is selected as the best path. Let the best $B$ children of path 4 be nodes 2, 8 and 9, ordered with descending correlation to the residue. In Fig.~\ref{Fig:IterationAOMP}b, the best child 2 is directly appended to Path 4, without increasing the number of paths. Fig.~\ref{Fig:IterationAOMP}c depicts addition of the second child 8, after which there appear five paths on the tree. As tree size is limited to $P=4$, path 2, which has the maximum cost, is removed. Finally, we consider node 9 in Fig.~\ref{Fig:IterationAOMP}d. The resultant path has higher cost than the other four paths. Hence, it is not added to the tree.

\subsubsection{Equivalent Path Pruning \label{Sec:PathEq}}

Neglecting insertion of equivalent paths to the tree is also important to improve the search performance. For this purpose, we define a path equivalency notion that also covers paths with different lengths: Let $S_1^{l_1}$ and $S_2^{l_2}$ be two paths of lengths $l_1$ and $l_2$, respectively, where $l_1 \geq l_2$. Let's define $S_{p,1}^{l_2}$ as the partial path that consists of the first $l_2$ nodes of $S_1^{l_1}$, i.e. $S_{p,1}^{l_2} = s_1^1,s_1^2,...,s_1^{l_2}$. $S_1^{l_1}$ and $S_2^{l_2}$ are equivalent if and only if $S_{p,1}^{l_2}$ and $S_2^{l_2}$ share the same set of nodes. In this case, orthogonality of the residue to the selected support, ensures that $S_{p,1}^{l_2}$ and $S_2^{l_2}$ are equivalent. Consequently, insertion of $S_2^{l_2}$ into the tree is unnecessary, as $S_{p,1}^{l_2}$ has already been expanded in previous iterations.

\begin{figure}[!t]
\centerline{\includegraphics[width=0.75\textwidth]{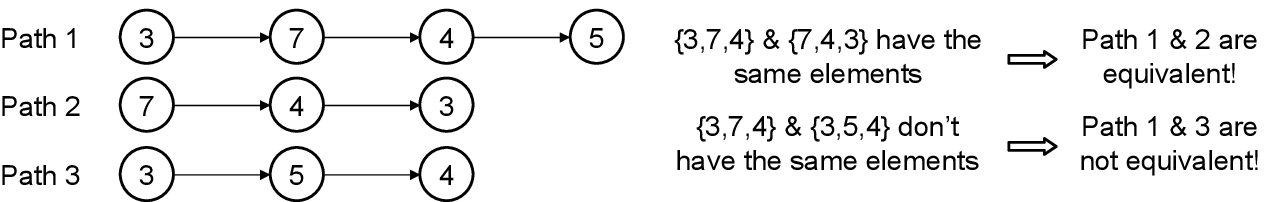}}
\caption{Path Equivalency: Path 1 and Path 2 are equivalent as first three nodes of Path 1 contain only nodes in Path 2. Path 3 is not equivalent to Path 1 as the node '5' is not an element of the first three nodes of Path 1. Note that orthogonal projections ensure Path 2 to select '5' as the next node, while there is no guarantee that Path 3 will select '7'.}
\label{Fig:PathEquivalency}
\end{figure}

Fig.~\ref{Fig:PathEquivalency} illustrates the path equivalency. Path 2 and the first three nodes of Path 1 share the same set of nodes, which makes Path 1 and Path 2 equivalent. Note that orthogonal projection ensures node 5 will be among the best children of path 2. On the contrary, Path 1 and Path 3 are not equivalent as the first three nodes of Path 1 and Path 3 are different. There exists no guarantee that node 7 will be among the best children of Path 3.

Let's now summarize extension of a selected partial path with these three pruning rules: First, the best $B$ children of the selected partial path $S$ are chosen as the dictionary atoms having highest inner-product with the residue. We obtain $B$ new candidate paths by appending $S$ one of these $B$ children. We apply Equivalent Path Pruning by eliminating candidates which are equivalent to already visited paths. For each remaining candidate, we first compute the residue via orthogonal projection of $\mathbf{y}$ onto $S$, and then the cost as discussed below. We remove $S$ from the tree and add the candidate paths. Finally, we prune the tree if number of paths exceeds $P$.

\subsection{Selection of the Most Promising Path \label{Sec:SelBestPath}}
A natural criterion for choosing the most promising path is the minimum residual error. Consequently, for a path $\mathbf{S^l}$ of length $l$, the evaluation function can be written as
\begin{equation}\label{Eq:SSRAStar1}
    g(\mathbf{S}^l) = \left\| \mathbf{r}^l \right\|_2 =  \left\| \mathbf{y} - \sum_{j=1}^lc^j\mathbf{s}^j \right\|_2.
\end{equation}
where $\mathbf{s}^j$ and $c^j$ denote the selected atom at stage j and the coefficient obtained after orthogonal projection of the residue onto the set of selected atoms, respectively.

As discussed in Section~\ref{Sec:AStar}, A* search employs an auxiliary function to compensate for different path lengths. The auxiliary function is important for comparing the multiple paths in the search tree. By proper evaluation of these paths, though any single one of them is limited to the RIP condition of OMP algorithm alone, A*OMP can relax the RIP condition, increasing the probability of finding a final path that is not altered by the linear dependency of the atoms in the dictionary. Ideally, the auxiliary function should mimic the decay of the residue along a path, which is impossible in practice. Below, we suggest three different methods which exploit different assumptions about the residue.

\subsubsection{Additive Cost Model \label{Sec:AddCost}}

The additive cost model assumes that the $K$ vectors in the representation make on average equal contributions to $\|\mathbf{y}\|_2$. That is, we assume that the average contribution of a vector is $\delta_e = \left\| \mathbf{y} \right\|_2/K$. Then, the unopened $K-l$ nodes of a partial path of length $l$ are expected to reduce $\|\mathbf{r}\|_2$ by $(K-l)\delta_e$. Combining this with (\ref{Eq:AStar1}), the auxiliary function should satisfy
\begin{equation}\label{Eq:SSRAStar2_2}
d_{add}(\mathbf{S}^l) \geq(K-l) \frac{\left\| \mathbf{y} \right\|_2}{K}.
\end{equation}

Consequently, we define the additive auxiliary function as
\begin{equation}\label{Eq:SSRAStar3}
    d_{add}(\mathbf{S}^l) \triangleq \beta (K-l) \frac{\left\| \mathbf{y} \right\|_2}{K} ,
\end{equation}
where $\beta$ is a constant greater than $1$. Finally, we obtain the additive cost function as
\begin{equation}\label{Eq:SSRAStar4}
    F_{add}(\mathbf{S}^l) = \left\| \mathbf{r}^l \right\|_2 - \beta \frac{(K-l)}{K} \left\| \mathbf{y} \right\|_2 .
\end{equation}
Here, $\beta$ acts as a regularization constant. If it is large, shorter paths are favored, making the search expand more candidates. When it becomes smaller, the search prefers longer paths. Note that favoring shorter paths increases the number of paths opened throughout the search, which improves the search at the expense of increased complexity. Hence, beta should be chosen to balance the available computational power or time restrictions and the recovery performance.

Note that $\delta_e = \left\| \mathbf{y} \right\|_2/K$ does not hold in general. However, (\ref{Eq:SSRAStar2_2}) requires this assumption only on average. Moreover, we intuitively expect the search to miss mostly the vectors with smaller contributions to $\|\mathbf{y}\|_2$, and for these, the additive auxiliary function satisfies (\ref{Eq:SSRAStar2_2}) with higher probabilities.

\subsubsection{Adaptive Cost Model \label{Sec:AdapCost}}

The auxiliary function can also be chosen adaptively by modifying the expectation on average contribution of an unopened node as:
\begin{equation}\label{Eq:SSRAStar5_1}
    \delta_e = \left(\left\|\mathbf{r}_{i}^{l-1}\right\|_2 - \left\|\mathbf{r}_{i}^{l}\right\|_2\right).
\end{equation}
Then, the adaptive auxiliary function should fulfill
\begin{equation}\label{Eq:SSRAStar5_2}
    d_{adap}(\mathbf{S}_i^l) \geq (K-l)(\left\|\mathbf{r}_{i}^{l-1}\right\|_2 - \left\|\mathbf{r}_{i}^{l}\right\|_2),
\end{equation}
where the subscript $i$ indicates the dependency on the particular path $\mathbf{S}_i^l$. (\ref{Eq:SSRAStar5_2}) can be justified by the fact that A* is configured to select first the vectors with higher contributions to $\mathbf{y}$. Hence, the residue is expected to decrease slower in later nodes than the initial nodes of a path.

As for the additive case, we incorporate $\beta>1$ to finally obtain the adaptive auxiliary function
\begin{equation}\label{Eq:SSRAStar5_3}
    d_{adap}(\mathbf{S}_i^l) = \beta(\left\|\mathbf{r}_{i}^{l-1}\right\|_2 - \left\|\mathbf{r}_{i}^{l}\right\|_2)(K-l).
\end{equation}
The adaptive cost function can then be written as follows:
\begin{equation}\label{Eq:SSRAStar5_4}
    F_{adap}(\mathbf{S}_i^l) = \left\| \mathbf{r}_i^l\right\|_2 - \beta(\left\|\mathbf{r}_{i}^{l-1}\right\|_2 - \left\|\mathbf{r}_{i}^{l}\right\|_2)(K-l),
\end{equation}
where the role of the regularization constant $\beta$ is very similar to the additive case.

\subsubsection{Multiplicative Cost Model \label{Sec:MulCost}}

In contrast to addition of the auxiliary function, multiplicative cost model path employs a weighting function. Here, we assume that each node reduces $\|\mathbf{r}\|_2$ by a constant ratio, $\alpha$. The multiplicative cost function is defined as
\begin{equation}\label{Eq:SSRAStar6}
    F_{mul}(\mathbf{S}_i^l) = \alpha^{K-l} g(\mathbf{S}_i^l)= \alpha^{K-l} \left\| \mathbf{r}_i^l \right\|_2.
\end{equation}
where $\alpha$ should be chosen between $0$ and $1$. The role of $\alpha$ is very close to that of $\beta$ for the additive cost function. When $\alpha$ is close to $0$, short paths are assigned very small costs, making the search to prefer them. On the contrary, if we choose $\alpha$ close to $1$, weighting is hardly effective on the cost function, hence longer paths will be favored.

In contrast to the additive one, adaptive and multiplicative cost models adjust the expected decay in $\mathbf{r}_i^l$ dynamically throughout the search. These dynamic structures are expected to provide a better modeling of the decrease in $\mathbf{r}_i^l$. In fact, the simulation results in Section~\ref{Sec:Results} clearly indicate that they improve the reconstruction accuracy.

\subsection{A* Orthogonal Matching Pursuit \label{Sec:A*OMP}}

We can now outline A*OMP: $I$ out of the $P$ paths, which are kept in a stack, are initialized as the $I$ vectors which best match $\mathbf{y}$ and the remaining $P-I$ paths are left empty. The cost for the empty paths is $\|\mathbf{y}\|_2$, hence they will be removed first. In each iteration, first, we select the path with minimum cost. We, then, expand the best $B$ children of the selected path applying the pruning rules discussed in Section~\ref{Sec:ExpSelPath}. Iterations are run until the selected path has length $K$. The pseudo-code for the algorithm is given in Algorithm~1. \ref{Alg:A*OMP}

\begin{algorithm}[!t]
\caption{ A* ORTHOGONAL MATCHING PURSUIT}
\label{Alg:A*OMP}
\begin{algorithmic}[]
\StartDefine
\State $P :=$ Maximum number of search paths
\State $I :=$ Number of initial search paths
\State $B :=$ Number of extended branches per iteration
\State $\mathbf{S}_i=\{\mathbf{s}^l_i\}$, matrix of atoms $\mathbf{s}^l_i$ on the $i$'th path
\State $\mathbf{c}_i = \{c^l_i\}$, vector of coefficients for the atoms on the $i$'th path
\State $L_i :=$  length of the $i$'th path
\State $C_i :=$ cost for selecting the $i$'th path
\EndDefine
\StartInit
\State $\mathbf{T} \gets \emptyset$
\For{i}{1}{I}\Comment{$I$ paths of length $1$}
    \State $\hat{n} \gets \argmax\limits_{n,\mathbf{v}_n \in \mathbf{\Phi}\setminus \mathbf{T}} \langle\mathbf{y},\mathbf{v}_n\rangle $
    \State $\mathbf{T} \gets \mathbf{T} \cup \mathbf{v}_{\hat{n}}$
    \State $\mathbf{s}^1_i \gets \mathbf{v}_{\hat{n}}$, $c^1_i \gets \langle\mathbf{y},\mathbf{v}_{\hat{n}}\rangle$
    \State $\mathbf{r}_i \gets \mathbf{y} -  c^1_i\mathbf{s}^1_i$
    \State $C_i = F(\mathbf{S}_i)$, $L_i = 1$
\EndFor
    \State $C_i = \|\mathbf{y}\|_2$, $L_i = 0$, $\forall i=I+1,I+2,...,P$
\State $best\_path \gets 1$
\EndInit
\While{$L_{best\_path} \neq K $}
    \State $\hat{p} \gets best\_path$ \Comment{first to replace}
    \State $\mathbf{T} \gets \mathbf{S}_{best\_path}$
    \For{i}{1}{B} \Comment{extensions per path pruning}
        \State $\hat{n} \gets \argmax\limits_{n,\mathbf{v}_n \in \mathbf{\Phi} \setminus \mathbf{T}} \langle\mathbf{r}_{best\_path},\mathbf{v}_n\rangle$
        \State $\mathbf{T} \gets \mathbf{T} \cup \mathbf{v}_{\hat{n}}$
        \State $\hat{\mathbf{S}} \gets \mathbf{S}_{best\_path} \cup \mathbf{v}_{\hat{n}}$ \Comment{candidate path}
        \State $\hat{\mathbf{c}} \gets \argmin\limits_{\alpha} \| \mathbf{y}- \mathbf{\hat{S}} \alpha  \|_{2}$ \Comment{Orthogonal projection}
        \State $\hat{C} \gets F(\hat{\mathbf{S}})$  \Comment{Cost of the candidate path}
         \If{$(\hat{C} < F(\mathbf{S}_{\hat{p}}))$ $\&$  \Comment{tree size pruning} \\
         $(\hat{\mathbf{S}} \neq {\mathbf{S}_j}$, $\forall j=1,2,...,P)$} \Comment{path equivalency}
            \State $\mathbf{S}_{\hat{p}} \gets \hat{\mathbf{S}}$, $\mathbf{c}_{\hat{p}} \gets \hat{\mathbf{c}}$, $C_{\hat{p}} \gets \hat{C}$
            \State $L_{\hat{p}} \gets L_{best\_path}+1$
            \State $\mathbf{r}_{\hat{p}} \gets \mathbf{y} -  \mathbf{S}_{\hat{p}}\mathbf{c}_{\hat{p}}$
            \State $\hat{p} \gets \argmax\limits_{i \in 1,2,...,P} C_i$ \Comment{to be replaced next}
        \EndIf
    \EndFor
    \State $best\_path \gets \argmin\limits_{i \in 1,2,...,P} C_i$ \Comment{select best path}
\EndWhile
\State \textbf{return} $\mathbf{S}_{best\_path}$, $\mathbf{c}_{best\_path}$
\end{algorithmic}
\end{algorithm}

We note that other termination criteria are also possible, including, for example, norm of the residue falling below a threshold, or no further reduction of the residue obtained.

\subsection{Complexity vs. Accuracy \label{Sec:CompVsAcc}}

The complexity of A*OMP approach arises from two points: The number of inner-product checks between the residue and dictionary atoms, and the number of orthogonal projections. The number of inner-product checks is equal to the number of iterations. Orthogonal projection, on the other hand, is necessary for each path, except the ones that are pruned by the equivalent path pruning. Hence, the number of these is equal to $B$ times the number of iterations minus the number of equivalent paths detected. Consequently, the important factors that govern the complexity of A*OMP are, first, the number of iterations and, second, the number of equivalent paths detected. However, it is not possible to find reasonable approximations of these. The only approximation to the number of paths is the upper bound that assumes opening of every possible node on the tree, which is obviously far away from being realistic. In order to give an insight on these, we investigate these experimentally in section~\ref{Sec:Results_DifDist}.

The pruning strategies of Section~\ref{Sec:ExpSelPath} can be seen as a trade-off between the accuracy and complexity of A*OMP. If we set $I=N$, $B=N$ and $P=\infty$, the algorithm will perform an exhaustive search, which is prohibitively complex. On the other hand, setting $I=1$ and $B=1$ yields OMP. A choice between the accuracy and complexity of the search can be adjusted by the pruning parameters. The accuracy is expected to increase with increasing these parameters, as demonstrated in section~\ref{Sec:Results_SearchParam}. In practice, these parameters, of course, may not be increased after some point, and regarding the results in section~\ref{Sec:Results_SearchParam}, it is also questionable if they will improve the performance after some point.

The cost model is also extremely important in the complexity-accuracy trade-off. An appropriate modeling of the decay in the residue improves the ability to predict branches on which the solution might lie. Therefore, the auxiliary function is important for both choosing the best path and pruning. With an appropriate choice, the trade-off between the complexity and accuracy is boosted in favor of accuracy, such as the dynamic cost functions improving the reconstruction ability in the first example in section~\ref{Sec:Results}. In addition, the auxiliary function parameters $\alpha$ and $\beta$ also affect the complexity-accuracy trade-off. Choosing $\beta \gg 1$ or $0<\alpha \ll 1$ makes the search favor shorter paths, leading to improvements in accuracy with longer search times. On the contrary, when $\beta$ and $\alpha$ are close to 1, the algorithm performs similar to OMP. These improvements are, of course, also expected to have some limits, for example, decreasing $\alpha$ does not improve the performance after some point, as demonstrated in section~\ref{Sec:Results_SearchParam}.

In order to get the best out of the search parameters, they should better be considered together. For example, reducing $\alpha$ increases the number of paths opened throughout the search. Consequently, a lower $\alpha$  value should be accompanied by an increment in the beam width $P$ in order to obtain better reconstruction results. This also holds when $\beta$ or $B$ is increased, which similarly increases the number of paths involved in the search. Examples in section~\ref{Sec:Results_SearchParam} illustrate this issue.

\section{Simulation Results \label{Sec:Results}}

We demonstrate sparse recovery via A*OMP in two problems in comparison to BP, SP and OMP. First of them is the recovery of a synthetically generated 1D signals, while the latter involves an image reconstruction problem. The simulations for A*OMP were performed using the AStarOMP software developed by the authors. The AStarOMP software incorporates a trie structure to implement the A* search tree in an efficient way. The orthogonalization over the residue is solved using the QR factorization. This software, and its MATLAB version, are available at http://myweb.sabanciuniv.edu/karahanoglu/research/.

\subsection{Reconstruction of Synthetically Generated 1D Data}

In this section, we evaluate three versions of A*OMP using additive, adaptive and multiplicative cost models. These are abbreviated as Add-A*OMP, Adap-A*OMP and Mul-A*OMP, respectively. The experiments cover different non-zero coefficient distributions, including uniform and Gaussian distributions as well as binary nonzero coefficients. We investigate reconstruction via Gaussian and Bernoulli observation matrices and compare different A*OMP parameters. Finally, we demonstrate A*OMP for reconstruction from noisy observations.

All the simulations in this section were repeated over $500$ randomly generated $K$-sparse samples of length $N=256$ from which $M=100$ random observations were taken via the observation matrix $\mathbf{\Phi}$. Reconstruction accuracy are given in terms of both the exact reconstruction rate and the average normalized mean squared error (NMSE), which is defined as the average ratio of the $\ell_2$ norm of the reconstruction error to $\|\mathbf{x}\|_2$ over the 500 test samples. For the noisy scenarios, we give the reconstruction error in the decibel scale, which we call the distortion ratio. Unless given explicitly, the following are common in all simulations: A*OMP parameters were set as $I=3$, $B=2$, $P=200$, $\beta = 1.25$ and $\alpha=0.8$. For each test sample, we employed an individual observation matrix $\mathbf{\Phi}$ whose entries were drawn from the Gaussian distribution with mean $0$ and standard deviation $1/N$.

\subsubsection{Different Coefficient Distributions \label{Sec:Results_DifDist}}

\begin{figure}[!t]
\centerline{\includegraphics[width=0.75\textwidth]{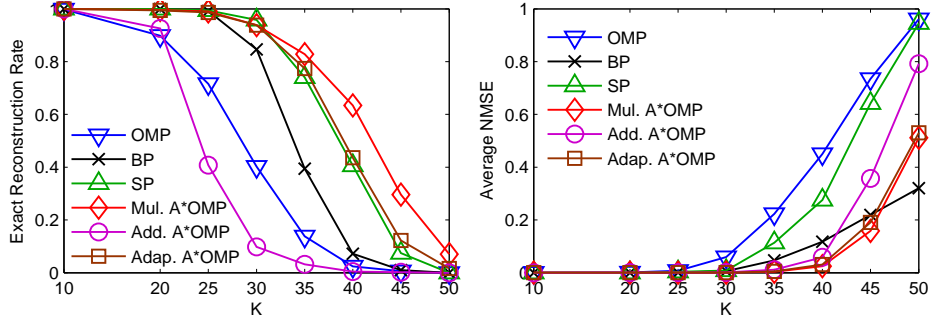}}
\caption{Reconstruction results over sparsity for uniform sparse signals employing Gaussian observation matrices.}
\label{Fig:results_toy_uniform}
\end{figure}

The first set of simulations employ sparse signals with nonzero coefficients drawn from the uniform distribution $U[-1,1]$. We refer to these signals as uniform sparse signals in the rest. The results of these simulations for $K$ from 10 to 50 are depicted in Fig.~\ref{Fig:results_toy_uniform}. In this test, Adap-A*OMP and Mul-A*OMP clearly provide lower average NMSE than BP, SP and OMP, except for $K=50$ where BP provides lower error. As expected, the average NMSE of OMP is the worst, while that of SP is only slightly better. BP provides lower error than SP and OMP, however it is still worse than A*OMP except for $K=50$. Even the Add-A*OMP, which employs no dynamic cost model, yields lower error than BP up to $K=40$. In addition to average NMSE, Mul-A*OMP, on general, yields higher exact recovery rates. Though SP yields high average NMSE, its exact recovery frequency competes with that of Mul-A*OMP up to $K=30$, and even exceeds it slightly at $K=30$.  For Add-A*OMP, the situation is contrary: Despite low average NMSE values, its exact reconstruction rate is even worse than that of OMP. These results indicate that the static cost model of Add-A*OMP most of the time fails at small nonzero coefficients. Adaptive and multiplicative cost models, which dynamically adjust the expected decay in $\|\mathbf{r}\|_2$ individually for each path, are clearly more effective for compensating path length differences.

\begin{figure}[!t]
\centerline{\includegraphics[width=0.75\textwidth]{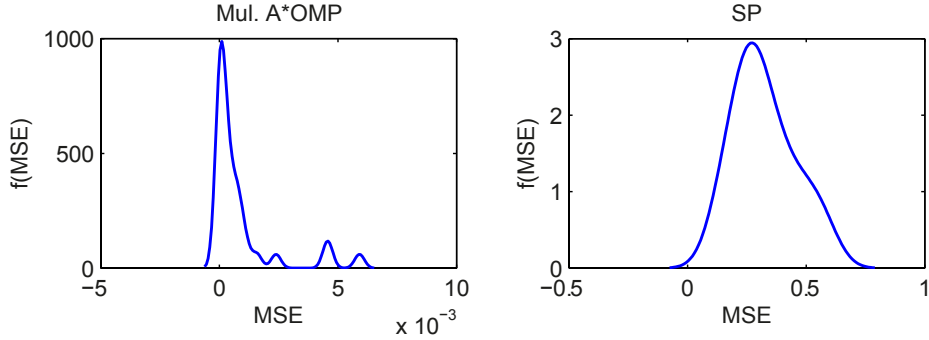}}
\caption{Probability density estimates of the NMSE for $K=30$.}
\label{Fig:results_errdist}
\end{figure}

\begin{figure}[!t]
\centerline{\includegraphics[width=0.75\textwidth]{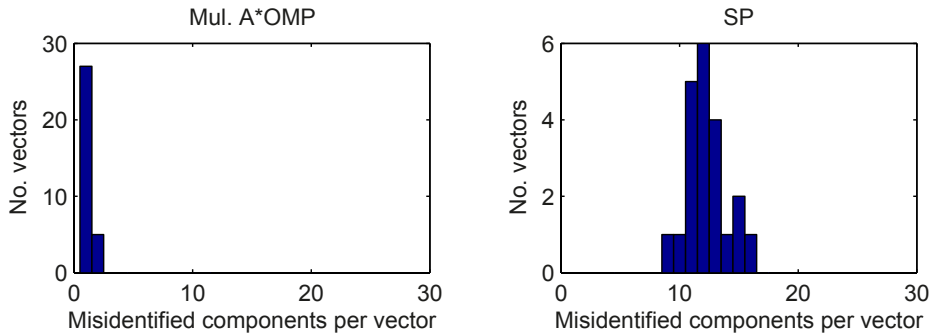}}
\caption{Number of misidentified entries per test sample for $K=30$.}
\label{Fig:results_errHist}
\end{figure}

As for SP, the exact recovery rate is much better than the NMSE suggests. This indicates that the amount of error SP makes per failure is much higher than that of the A*OMP algorithm. To visualize this fact, the probability density estimates of the error are depicted in Fig.~\ref{Fig:results_errdist} for SP and Mul-A*OMP. These were computed using Gaussian kernels over NMSE of the test vectors which could not be exactly reconstructed for $K=30$. The figures state that NMSE values on the order of $10^{-3}$'s for Mul-A*OMP, while for SP, they range up to 0.8, with mean about 0.3. This arises from the difference in the average number of misidentified elements per failure, which is shown in Fig.~\ref{Fig:results_errHist} for $K=30$. Mul-A*OMP has misidentified only one or two of the 30 components, while SP has missed 9 to 16 components, and on average about 12 per failure. These figures indicate that if the reconstruction is not exact, SP almost completely fails, however A*OMP can still reconstruct the desired vector with small amount of error, which is less than 1\% of the signal norm for K = 30.

As discussed in section~\ref{Sec:CompVsAcc}, the two important factors for the complexity of A*OMP are the average A*OMP iterations per vector and the average equivalent paths detected per vector. Table~\ref{Table:NoIter_toy_uniform} states the average A*OMP iterations per vector in this scenario in comparison to the upper bound on the number of A*OMP iterations. This upper bound can easily be obtained as $I\cdot(2^{K-1}-1)$ for $B=2$ by assuming that all of the opened partial paths are selected one by one as the best path throughout the search. The actual number of iterations is incomparably lower than this upper bound. Moreover, though the upper bound increases exponentially with $K$, the actual number of iterations exhibit a much lower slope. The second important factor, the average number of equivalent paths per vector is given in Table~\ref{Table:NoEqPaths_toy_uniform}. These numbers are comparable to the number of iterations, which states the effectiveness of the equivalent path pruning rule. These results indicate that pruning and proper selection of the cost model make it possible to run the search for cases where the upper bound becomes unpractically high.

\renewcommand{\arraystretch}{1.1}
\begin{table}[!h]
\centering
\caption{Average A*OMP iterations per vector for uniform sparse signals}
{\footnotesize{
\begin{tabular}{p{2cm} x{1.25cm} x{1.25cm}  x{1.25cm} x{1.25cm}}  \hline \hline
& \multicolumn{4}{c}{$K$} \tabularnewline \cline{2-5}
&10&20&30&40\tabularnewline \hline \hline
Mul-A*OMP&13.8& 164  &1695 & 4177\tabularnewline \hline
Adap-A*OMP&19& 167.4 &2443 & 6109 \tabularnewline \hline
Upper Bound & 1533 & $1.57\cdot10^6$ & $1.61\cdot10^9$ & $1.65\cdot10^{12}$ \tabularnewline \hline \hline
\end{tabular}}}
\label{Table:NoIter_toy_uniform}
\end{table}

\renewcommand{\arraystretch}{1.1}
\begin{table}[!h]
\centering
\caption{Average equivalent paths per vector for uniform sparse signals}
{\footnotesize{
\begin{tabular}{p{2cm} x{1cm} x{1cm}  x{1cm} x{1cm}}  \hline \hline
& \multicolumn{4}{c}{$K$} \tabularnewline \cline{2-5}
&10&20&30&40\tabularnewline \hline \hline
Mul-A*OMP  & 4.4 & 114.1 & 975.2 & 1776  \tabularnewline \hline
Adap-A*OMP & 11.2 & 126.6 & 1355 & 1831 \tabularnewline \hline \hline
\end{tabular}}}
\label{Table:NoEqPaths_toy_uniform}
\end{table}

Finally, in order to provide an insight about the speed of the search, we list in Table~\ref{Table:Time_toy_uniform} the average run-times for Mul-A*OMP, Adap-A*OMP and OMP on a modest Pentium Dual-Core CPU at 2.3GHz. These were obtained using the AStarOMP software and a similar OMP implementation developed by the authors specially for obtaining comparable run-times. Note that the structure of A*OMP makes it possible to process the $B$ candidates in parallel at each iteration. Moreover, the search can easily be modified to open more than one promising path per iteration in parallel. Hence, these run-times can be significantly reduced by parallel programming, which is beyond the scope of this paper.

\renewcommand{\arraystretch}{1.1}
\begin{table}[!h]
\centering
\caption{Average run-time in sec. per vector for uniform sparse signals}
{\footnotesize{
\begin{tabular}{p{2cm} x{1cm} x{1cm}  x{1cm} x{1cm}}  \hline \hline
& \multicolumn{4}{c}{$K$} \tabularnewline \cline{2-5}
&10&20&30&40\tabularnewline \hline \hline
OMP  &0.0012 & 0.0025 & 0.0036 & 0.0050 \tabularnewline \hline
Mul-A*OMP  & 0.0022 & 0.0261 & 0.3158 & 0.8292 \tabularnewline \hline
Adap-A*OMP & 0.0032 & 0.0276 & 0.4601 & 1.1525 \tabularnewline \hline \hline
\end{tabular}}}
\label{Table:Time_toy_uniform}
\end{table}

\begin{figure}[!t]
\centerline{\includegraphics[width=0.75\textwidth]{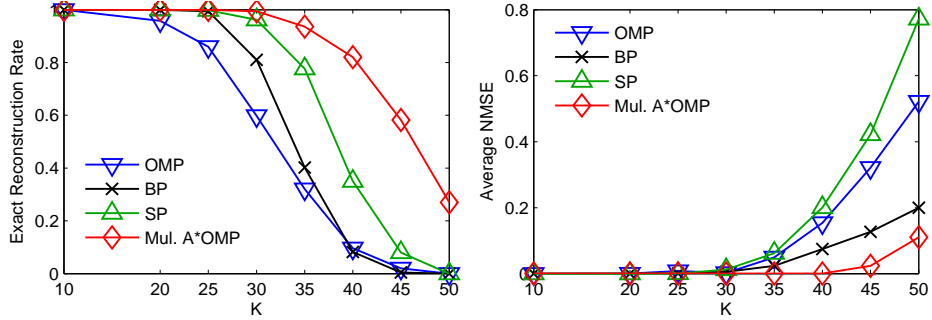}}
\caption{Reconstruction results over sparsity for Gaussian sparse vectors using Gaussian observation matrices.}
\label{Fig:results_toy_Gauss}
\end{figure}

For the second set of simulations, we employ Gaussian sparse vectors, whose nonzero entries were drawn from the standard Gaussian distribution. Fig.~\ref{Fig:results_toy_Gauss} depicts the average NMSE and exact reconstruction rates for this test. In this scenario, Mul-A*OMP provides clearly better reconstruction than BP, SP and OMP. We observe that it provides both lower NMSE and higher exact reconstruction rate than all the other algorithms. SP yields the second best exact reconstruction rate, however, its average NMSE is the worst, as a consequence of the almost complete failure of a non-exact reconstruction.

\begin{figure}[!t]
\centerline{\includegraphics[width=0.75\textwidth]{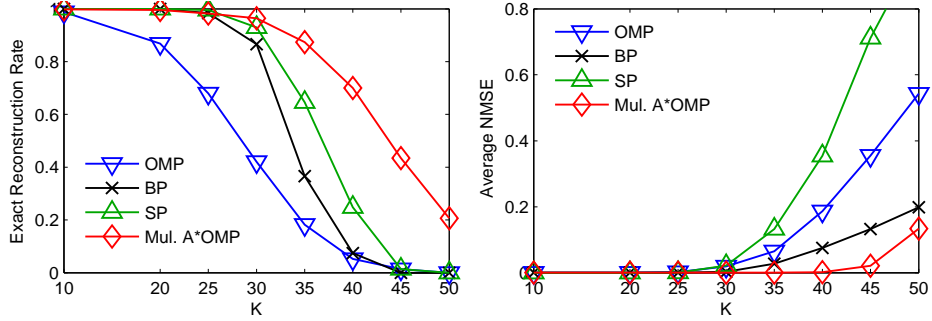}}
\caption{Reconstruction results over sparsity for Gaussian sparse vectors using Bernoulli observation matrices.}
\label{Fig:results_Gauss_bernoulli}
\end{figure}

In order to question the choice of the observation matrix, we repeat the last scenario with observation matrices drawn from the Bernoulli distribution. The average NMSE and exact reconstruction rates for this test are illustrated in Fig.~\ref{Fig:results_Gauss_bernoulli}. Comparing Fig.~\ref{Fig:results_Gauss_bernoulli} with Fig.~\ref{Fig:results_toy_Gauss}, we observe that the average NMSE values remain quite unaltered for Mul-A*OMP and BP, while that for SP increases. Mul-A*OMP leads to the least amount of error. As for exact reconstruction, only BP keeps the same rates, while the rates of all others fall. BP and SP compete with Mul-A*OMP until $K=25$, where SP is slightly better. When $K$ further increases, Mul-A*OMP has the highest exact recovery frequency.

\begin{figure}[!t]
\centerline{\includegraphics[width=0.75\textwidth]{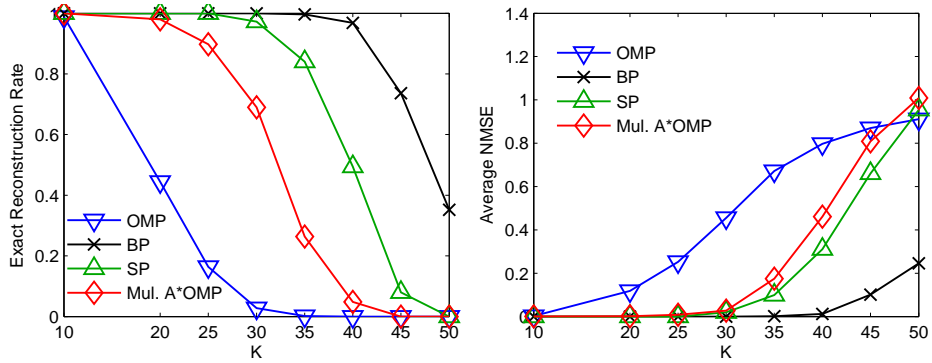}}
\caption{Reconstruction results over sparsity for sparse binary signals using Gaussian observation matrices.}
\label{Fig:results_toy_binary}
\end{figure}

Next problem is the reconstruction of sparse binary vectors, where the nonzero coefficients were selected as 1. The results are shown in Fig.~\ref{Fig:results_toy_binary}. We observe that BP clearly yields better reconstruction than the others in this case. SP also performs better than A*OMP. The failure of A*OMP is related to the fact that this is a particularly challenging case for OMP-type of algorithms \cite{Dai:SP}. OMP is shown to have non-uniform guarantees, and, though mathematical justification of A*OMP is quite hard, this non-uniformity seems to be carried over to A*OMP for this type of signals. In contrast, for sparse binary signals, $\ell_0$ norm of the correct solution is exactly equal to its $\ell_1$ norm, which might be considered as an advantage for BP in this particular scenario. The results of this scenario, however, should not be very discouraging since sparse binary vectors represent a limited subset of the real world problems.

\subsubsection{Performance over Different Observation Lengths}

\begin{figure}[!t]
\centerline{\includegraphics[width=0.75\textwidth]{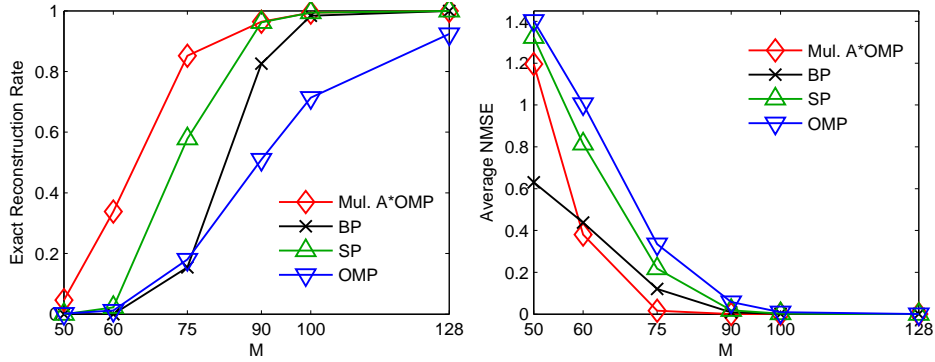}}
\caption{Reconstruction results over observation lengths for uniform sparse signals where $K=25$ using a single Gaussian observation matrix for each $M$.}
\label{Fig:results_uniform_M}
\end{figure}

Another interesting test case is the reconstruction ability when the observation length, $M$, changes. Fig.~\ref{Fig:results_uniform_M} depicts the recovery performance over $M$ for uniform sparse signals where $K=25$. For each $M$ value, a single Gaussian observation matrix is employed to obtain observations from all signals. We observe that Mul-A*OMP is the best in terms of the exact recovery rates, while SP and BP compete it for $M\geq90$ and $M\geq100$, respectively. The average NMSE of Mul-A*OMP is also lower than the others except for the case of $M=50$ where BP provides lower error than Mul-A*OMP.

\subsubsection{Comparison of Different Search Parameters \label{Sec:Results_SearchParam}}

Choosing the search parameters is an important issue for the A*OMP algorithm. This was discussed above in section~\ref{Sec:CompVsAcc}, indicating two main points: The reconstruction performance of the search might be increased by modifying the search parameters to explore more paths in the search at the expense of increased iterations and search times. In order to demonstrate this, we consider two scenarios. First, we vary $\alpha$, and later $B$ together with $P$.

\begin{figure}[!t]
\centerline{\includegraphics[width=0.75\textwidth]{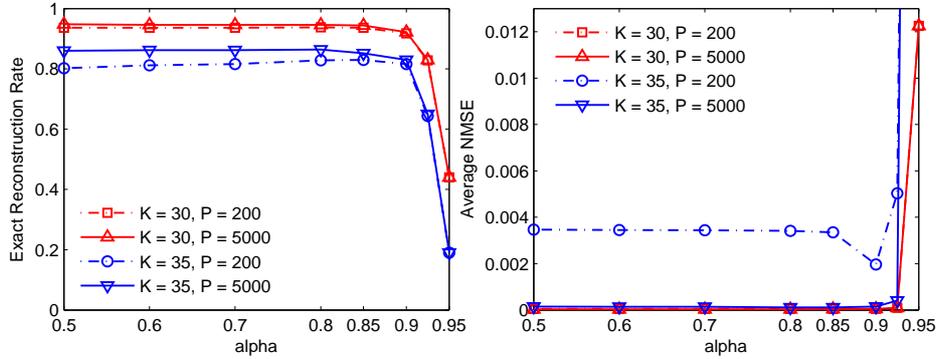}}
\caption{Reconstruction results over $\alpha$ for uniform sparse signals using Gaussian observation matrices.}
\label{Fig:results_uniform_alphaP}
\end{figure}

Fig.~\ref{Fig:results_uniform_alphaP} depicts the performance of Mul-A*OMP over $\alpha$ for uniform sparse signals with $K=30$ and $K=35$. The dashed and solid lines indicate results for $P=200$ and $P=5000$, respectively. For $K=30$, the reconstruction performance increases when $\alpha$ is reduced from 0.95 to about 0.8, whereas any further reduction of $\alpha$ does not significantly affect the performance. In addition, there is hardly any difference between selecting $P=200$ and $P=5000$. This suggests that setting $P=200$ and $\alpha\approx0.8$  seems to be enough for $K=30$. When $K=35$, however, more paths are involved in the search, and increasing $P$ improves the reconstruction. When $P=200$, reducing $\alpha$ below 0.9 does not improve but slightly degrade the performance. On the contrary, if $P$ is increased to 5000, the reconstruction  is improved until $\alpha$ is reduced to $0.8$, below which the reconstruction performance does not change. Though not given in the figures, the authors have observed that setting $P>5000$ has hardly any effect on the reconstruction. These results demonstrate that reducing $\alpha$ improves the reconstruction until some convergence point. Table~\ref{Table:NoIter_uniform_alphaP} lists the average number of search iterations while $\alpha$ and $P$ are varied. We observe that decreasing $P$ and increasing $\alpha$ increase the number of paths involved, which clarifies complexity-accuracy trade off that leads to improved recovery performance at the expense of increased complexity.

\renewcommand{\arraystretch}{1.1}
\begin{table}[!t]
\centering
\caption{Average Mul-A*OMP iterations per vector  wrt. $\alpha$ and $P$ for uniform sparse signals with $K=35$}
{\footnotesize{
\begin{tabular}{x{1.2cm} x{1cm} x{0.9cm} x{0.9cm}  x{0.9cm} x{0.9cm}}  \hline \hline
&$\alpha=0.5$& $\alpha=0.6$ & $\alpha=0.7$ & $\alpha=0.8$ & $\alpha=0.9$ \tabularnewline \hline \hline
$P=200$  & 4158  & 3927  & 3565  & 2932 & 1353\tabularnewline \hline
$P=5000$ & 58204 & 51710 & 41781 & 25527 & 4026 \tabularnewline \hline \hline
\end{tabular}}}
\label{Table:NoIter_uniform_alphaP}
\end{table}

\begin{figure}[!t]
\centerline{\includegraphics[width=0.75\textwidth]{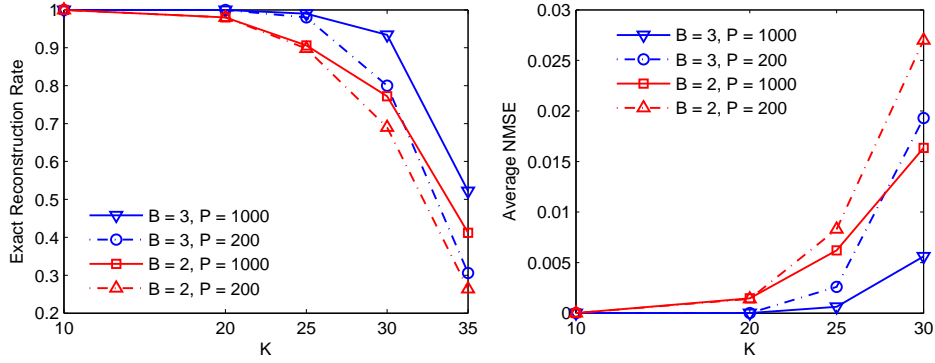}}
\caption{Reconstruction results for sparse binary signals for $B=2$ and $B=3$ using Gaussian observation matrices.}
\label{Fig:results_binary_B}
\end{figure}

Next, we illustrate the performance of Mul-A*OMP with $B=2$ and $B=3$ for sparse binary signals in Fig.~\ref{Fig:results_binary_B}. The experiment was repeated for $P=200$ and $P=1000$, which are depicted by dashed and solid lines, respectively. We observe that increasing $B$ from 2 to 3 improves the reconstruction. This improvement is further enhanced by increasing $P$ from 200 to 1000 when $K\geq25$, where a larger search stack can better cover for the increased number of paths involved in the search. Table~\ref{Table:NoIter_binary_B} lists the average number of search iterations, which increase with $B$ and $P$. Hence, the improvement, as above, is obtained at the expense of complexity.

\renewcommand{\arraystretch}{1.1}
\begin{table}[!t]
\centering
\caption{Average Mul-A*OMP iterations wrt. $B$ and $P$ per vector in the sparse binary problem}
{\footnotesize{
\begin{tabular}{p{1cm} x{0.001cm} x{0.9cm} x{0.001cm} x{0.9cm}  x{0.001cm} x{0.9cm}  x{0.001cm} x{0.9cm}}  \hline \hline
&& \multicolumn{3}{c}{P = 200} && \multicolumn{3}{c}{P = 1000} \tabularnewline \cline{3-5} \cline{7-9}
&&B=2&&B=3&&B=2&&B=3\tabularnewline \hline \hline
$K = 10$ && 48 && 114 && 48 && 114  \tabularnewline \hline
$K = 20$ && 1046 && 2095 && 1275 && 7159 \tabularnewline \hline
$K = 30$ && 3424 && 4249 && 12278 && 18240 \tabularnewline \hline \hline
\end{tabular}
}}
\label{Table:NoIter_binary_B}
\end{table}

The results in this section explain how the performance of A*OMP can be adjusted by the search parameters. The mechanism behind is simple: Increasing the number of paths explored by the search improves the results, until a convergence point, at the expense of increasing the complexity. According to the experimental results, one advantage is that even with modest settings such as $I=2$, $P=200$ and $\alpha=0.8$ employed in the experiments, A*OMP can provide high exact recovery frequencies and lower error than the other candidates for uniform and Gaussian sparse signals. This indicated that A*OMP recovery, at least in these cases, is quite robust against the choice of search parameters.

\subsubsection{Reconstruction from Noisy Observations}

\begin{figure}[!t]
\centerline{\includegraphics[width=0.75\textwidth]{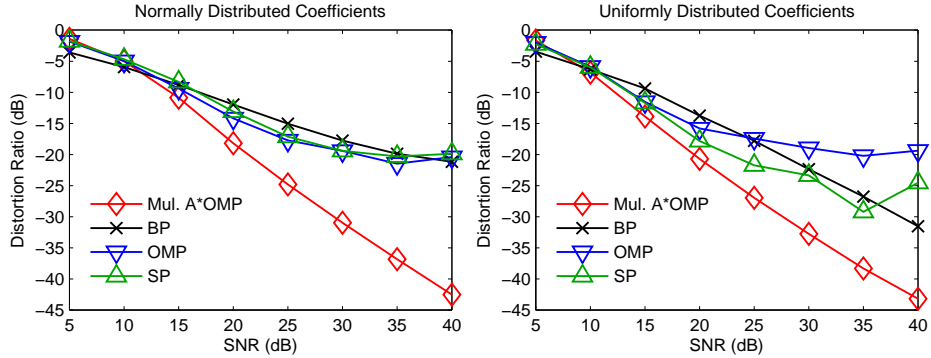}}
\caption{Average NMSE over SNR for reconstruction of sparse signals from noisy observations using Gaussian observation matrices.}
\label{Fig:results_noisy}
\end{figure}

Fig.~\ref{Fig:results_noisy} illustrates recovery results where the observation vectors are contaminated by white gaussian noise at different SNR levels. Here, $K$ is $25$ and $30$ for Gaussian and uniform sparse signals, respectively. The results are shown in terms of the distortion ratio in the decibel scale for better comparison. We observe that Mul. A*OMP produces less error than BP, SP and OMP for about 10dB and higher. When SNR decreases, BP starts to be more effective than the greedy algorithms.

\subsection{Reconstruction of Images}

We finally simulate the reconstruction ability of A*OMP on some commonly used $512 \times 512$ images including 'Lena', 'Tracy', 'cameraman', etc. The images were reconstructed in $8 \times 8$ blocks which provide important advantages that reduce the complexity and memory requirements of the search. First, without block-processing, the reconstruction problem requires searching among $N=512^2=262144$ dictionary atoms. However, block-processing reduces the problem to $4096$ subproblems with $N=64$, which is more efficient as each subproblem requires a search in 4096-fold reduced dimensionality. Second, block-processing reduces the total number of search paths drastically. To illustrate, let's set $B=2$. From Section~\ref{Sec:ExpSelPath}, the number of search paths for each $K$-sparse block is upper bounded by $I \cdot 2^{(K-1)}$. Then, for the whole image, the upper bound becomes $4096 \cdot I \cdot 2^{(K-1)}=I \cdot 2^{(K+11}$. If no block processing were involved, the upper bound would be $I \cdot 2^{D}$ where $D \gg K+11$. Finally, block-processing also reduces the length of the involved paths. Note that the block structure is shared by all involved recovery methods.

The simulations were performed with five $512 \times 512$ grayscale images using the 2D Haar Wavelet basis $\mathbf{\Psi}$. Note that in this case, the dictionary is not $\mathbf{\Phi}$, but the holographic basis $\mathbf{V}=\mathbf{\Phi}\mathbf{\Psi}$. Images were first preprocessed such that each $8 \times 8$ block is K-sparse in the 2D Haar Wavelet basis, where $K=14$.
A single observation matrix $\mathbf{\Phi}$ of size $ M  \times {N} $, which was randomly drawn from the Gaussian distribution with mean 0 and standard deviation $1/N$, was employed to compute the measurements of length $M=32$ from each block. Mul-A*OMP and Adap-A*OMP were run for both $B=2$ and $B=3$. We selected $I=3$ and $P=200$. The cost function parameters were set to $\alpha = 0.5$ and $\beta = 1.25$.

\renewcommand{\arraystretch}{1.1}
\begin{table}[!t]
\centering
\caption{PSNR values for images reconstructed using different algorithms}
{\footnotesize{
\begin{tabular}{p{1.3cm}  x{0.5cm}  x{0.65cm}  x{0.5cm}  x{0.55cm}  x{0.55cm}  x{0.55cm}  x{0.55cm}}  \hline \hline
&\multirow{2}{*}{BP}&\multirow{2}{*}{OMP}&\multirow{2}{*}{SP} & \multicolumn{2}{c}{Mul-A*OMP} & \multicolumn{2}{c}{Adap-A*OMP} \tabularnewline \cline{5-6} \cline{7-8}
&&&&B=2&B=3&B=2&B=3\tabularnewline \hline \hline
Lena      & 33.5 & 29.6 & 27.5 & 36.4 & 38.3 & 35.2 & 37 \tabularnewline \hline
Tracy     & 40.6 & 36.8 & 33.9 & 44.8 & 46.4 & 44.5 & 45.5 \tabularnewline \hline
Pirate    & 31.7 & 27.7 & 25.3 & 33.6 & 34.5 & 32.8 & 34.2 \tabularnewline \hline
Cameraman & 34.4 & 30.7 & 28.5 & 38.4 & 40.2 & 36.7 & 39.5 \tabularnewline \hline
Mandrill  & 28.3 & 24.4 & 22.1 & 30.3 & 31.3 & 29.3 & 30.8 \tabularnewline \hline \hline
\end{tabular}
}}
\label{Table:compImage}
\end{table}

\begin{figure*}[!t]
\centerline{\includegraphics{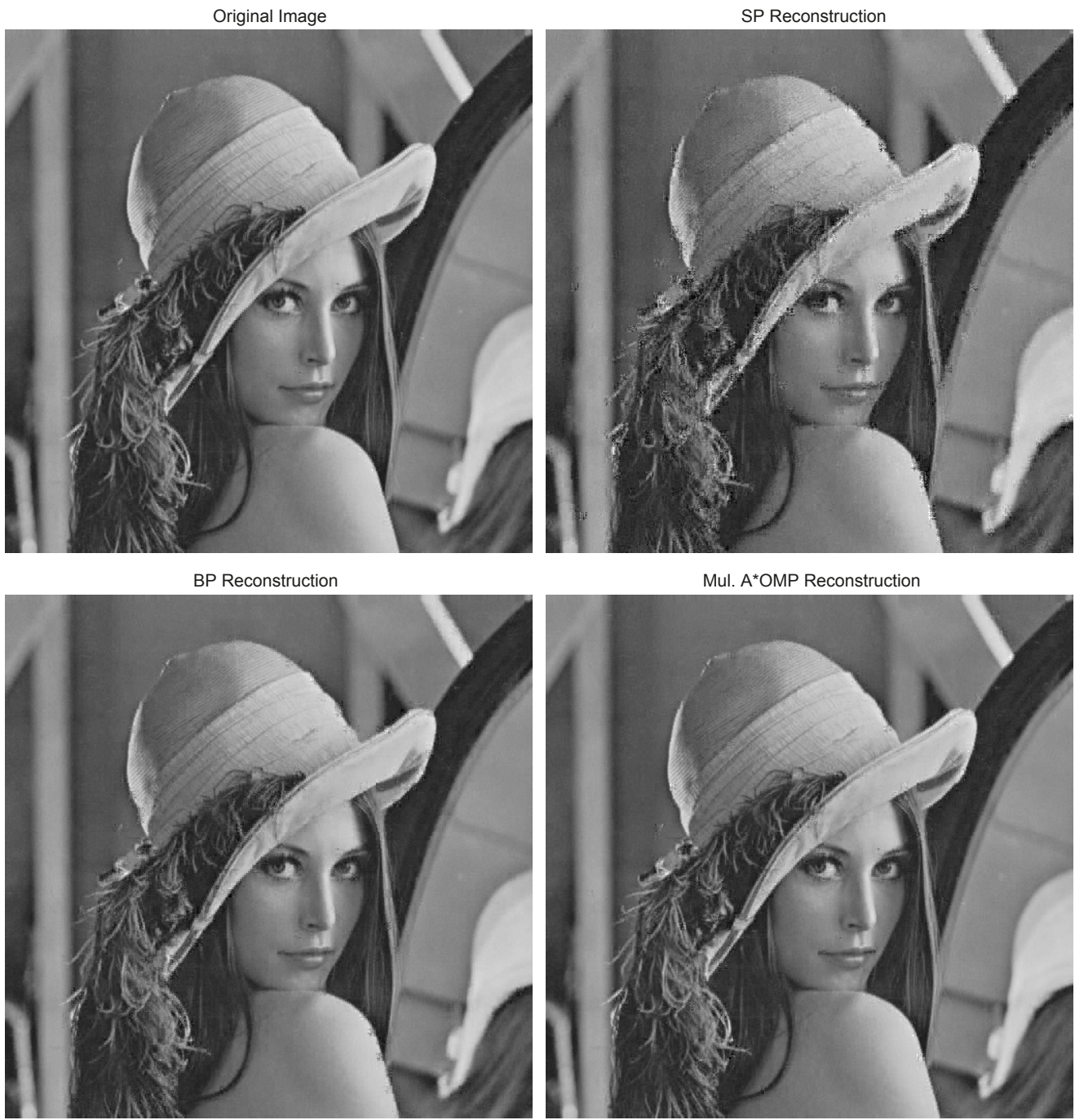}}
\caption{Reconstructions of image 'Lena' using different algorithms}
\label{Fig:lena}
\end{figure*}

\begin{figure}[!t]
\centerline{\includegraphics[width= 0.75\textwidth]{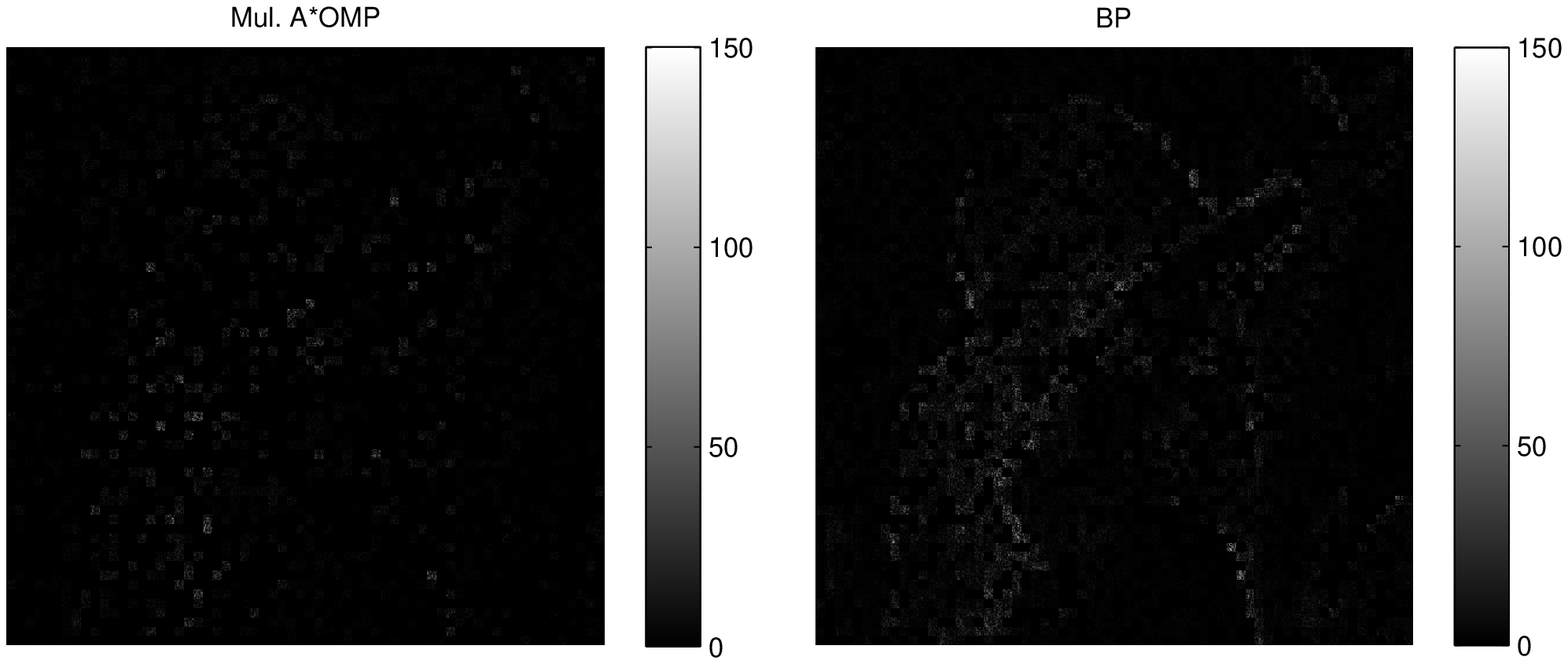}}
\caption{Reconstruction error per pixel of image 'Lena' for Mul-A*OMP with $B=3$ and BP.}
\label{Fig:lena_dif}
\end{figure}

Table~\ref{Table:compImage} lists the peak Signal-to-Noise ratio (PSNR) of reconstructed images. A*OMP yields better reconstruction than the other methods. Increasing $B$ from 2 to 3 further improves the reconstruction performance. A*OMP improves PSNR up to 5.8 dB, and 4.4 dB on average over BP. As an example, Fig.~\ref{Fig:lena} depicts reconstruction of 'lena' using SP, BP and Mul-A*OMP with $B=3$. Mul-A*OMP reconstruction provides lower error, which can be observed better in Fig.~\ref{Fig:lena_dif} illustrating the absolute error per pixel for BP and Adap-A*OMP reconstructions. For BP, errors are concentrated around boundaries and detailed regions, while Mul-A*OMP clearly produces less distortion all around the image.

\section{Conclusion}

This work introduces a novel CS reconstruction approach, A*OMP, which is based on an effective combination of OMP with A* search. This semi-greedy method performs a tree-search, that favors the paths minimizing the cost function on-the-fly. In order to compare paths with different lengths, novel dynamic cost functions, which show better reconstruction in the provided experiments, are defined. Pruning strategies are introduced to limit the search running times. A complexity-accuracy trade-off is provided via adjustment of the search parameters. In the provided experiments, A*OMP, with some modest settings, performs better reconstruction for uniform and Gaussian sparse signals, and for images than BP and SP. It also shows robust performance under presence of noise. BP and SP perform better than A*OMP for the sparse binary signals which constitute a limited subset of the real world problems. Moreover, as demonstrated, the A*OMP reconstruction in this case can be improved by modifying the search parameters at the expense of complexity.

To conclude, the demonstrated reconstruction performance of A*OMP indicates that it is a promising approach, that is capable of reducing the reconstruction errors significantly.

\bibliographystyle{model1-num-names}

\vskip5mm
\textbf{Nazim Burak Karahanoglu} received his B.S. degree in Electrical and Electronic Engineering from METU, Ankara  in 2003 and M.S. degree in Computational Engineering from the Friedrich-Alexander University of Erlangen-Nuremberg, Germany in 2006. He has been with the Information Technologies Institute of the Scientific and Technological Research Council of Turkey (TUBITAK) since 2008. He is also a Ph.D. student at the Electronics Engineering Department of Sabanci University, Turkey. His research interests include compressed sensing and sonar signal processing.
\vskip5mm
\textbf{Hakan Erdogan} is an assistant professor at Sabanci University in Istanbul, Turkey. He received his B.S. degree in Electrical Engineering and Mathematics in 1993 from METU, Ankara and his M.S. and Ph.D. degrees in Electrical Engineering: Systems from the University of Michigan, Ann Arbor in 1995 and 1999 respectively. He was with the Human Language Technologies group at IBM T.J. Watson Research Center, NY between 1999 and 2002. He has been with Sabanci University since 2002. His research interests are in developing and applying probabilistic methods and algorithms for multimedia information extraction.

\end{document}